\RequirePackage{fix-cm}
\documentclass[smallcondensed,smallextended]{svjour3}
\smartqed  
\usepackage{latexsym,amsmath,amssymb,amsfonts,mathbbol,graphicx,color}
\begin{document}

\title{Syndrome Measurement Strategies for the [[7,1,3]] Code
}


\author{Yaakov S. Weinstein}


\institute{Quantum Information Science Group, {\sc Mitre}, 200 Forrestal Rd., Princeton, NJ 08540, USA
              \email{weinstein@mitre.org}             \\
}

\date{Received: date / Accepted: date}

\maketitle

\begin{abstract}
Quantum error correction (QEC) entails the encoding of quantum information into a QEC code space, measuring error syndromes to properly locate and identify errors, and, if necessary, applying a proper recovery operation. Here we compare three syndrome measurement protocols for the [[7,1,3]] QEC code: Shor states, Steane states, and one ancilla qubit by simulating the implementation of 50 logical gates with the syndrome measurements interspersed between the gates at different intervals. We then compare the fidelities for the different syndrome measurement types. Our simulations show that the optimal syndrome measurement strategy is generally not to apply syndrome measurements after every gate but depends on the details of the error environment. Our simulations also allow a quantum computer programmer to weigh computational accuracy versus resource consumption (time and number of qubits) for a particular error environment. In addition, we show that applying syndrome measurements that are unnecessary from the standpoint of quantum fault tolerance may be helpful in achieving better accuracy or in lowering resource consumption. Finally, our simulations demonstrate that the single-qubit non-fault tolerant syndrome measurement strategy achieves comparable fidelity to those that are fault tolerant. 
\keywords{quantum error correction \and quantum fault tolerance \and syndrome measurements}
\PACS{03.67.Mn \and 03.67.Bg \and 03.67.Pp}
\end{abstract}

\section{Introduction}

Quantum error correction (QEC) codes can be used to make quantum information robust against errors \cite{book,ShorQEC,CSS,Steane}. Quantum error correction consists of three basic steps. First, the quantum information, some number of logical qubits, is encoded into a larger number of physical qubits. Second, syndrome measurements (SM), parity measurements between multiple qubits, are performed to determine the presence and location of an error. Finally, if an error is detected, a recovery operation is applied to correct the error.  

Quantum fault tolerance (QFT) \cite{Preskill,ShorQFT,G,AGP} is the framework that allows for successful quantum computation despite a finite probability of error in basic computational gates. Adherence to the framework requires that manipulation of the encoded quantum information, such as the performance of gates and measurements, be done within the QEC code. In addition, the tenents of QFT require that all manipulations be done in such a way that if an error does occur to a (physial) qubit it cannot spread to other qubits. QFT generally assumes that SM are performed after every logical gate so as to ensure no errors have occurred. However, SM are expensive both in time and number of qubits. Hence, recent studies have explored whether one can implement SM less often than after every gate \cite{How,How2,WIPK}. In this paper we expand on this work by exploring a variety of SM protocols and comparing the accuracy of implementation and resource cost of the different protocols. In addition, we show how deviating from the strict tenets of QFT, both by adding additional SM and using non-fault tolerant SM, may improve accuracy of save resources.     

\begin{figure}
\begin{center}
\includegraphics[width=10cm]{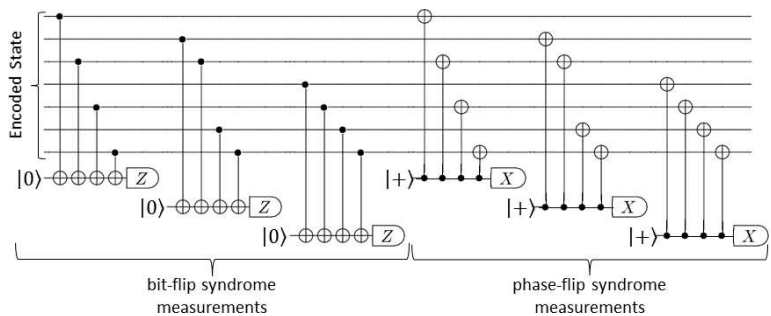}
\caption{Circuit for non-fault tolerant syndrome measurements for the [[7,1,3]] QEC code using single qubit ancilla. }
\label{1Q}
\end{center}
\end{figure}

A paradigmatic example of a QEC code is the Steane [[7,1,3]] code \cite{Steane} in which one logical qubit is encoded into 7 physical qubits. The encoding is robust against all single (physical) qubit errors. In the initial formulation of the code 6 SM were needed to detect and identify an error, 3 for bit-flip errors and 3 for phase-flip errors. Each SM required one ancilla qubit as shown in Fig.~\ref{1Q}.

Through the lens of QFT, however, the SM scheme of Fig.~\ref{1Q} is flawed. This is because errors on one qubit can spread uncontrollably throughout the circuit. For example, an error to one of the ancilla qubits utilized for the SM can easily affect the four `data' qubits it interacts with and from there spread to even more qubits. To implement SM that will adhere to the rules of QFT any ancilla qubit should interact with only one data qubit. For the [[7,1,3]] QEC code a number of such SM schemes have been developed. 

\begin{figure}
\begin{center}
\includegraphics[width=10cm]{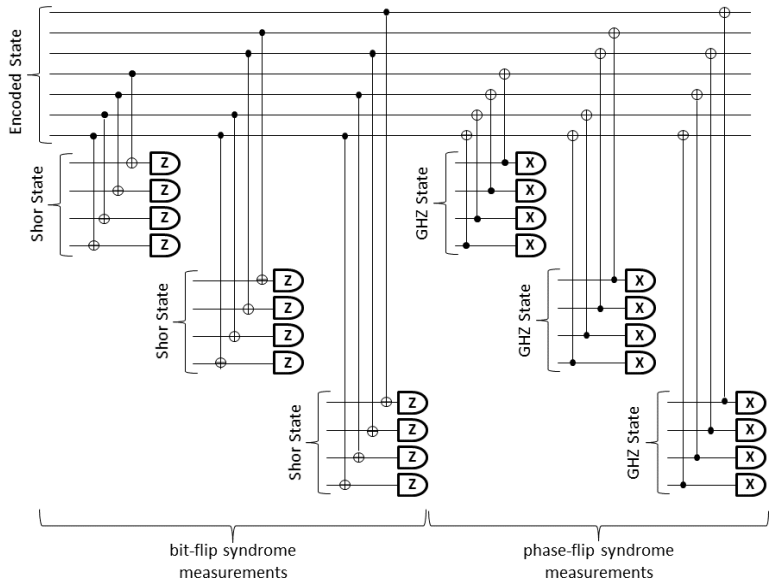}
\caption{Circuit for fault tolerant syndrome measurements for the [[7,1,3]] QEC code using Shor states. To ensure fault tolerance the set of six syndromes is repeated until the same syndrome is read out twice. }
\label{Shor}
\end{center}
\end{figure}

One method of performing the [[7,1,3]] code SM in a fault tolerant fashion is to substitute each single ancilla qubit of Fig.~\ref{1Q} with a four-qubit Shor state \cite{ShorQFT}. Shor states are GHZ states with Hadamard gates appended to each qubit. The Shor state is constructed in a fault tolerant manner by applying appropriate verifications \cite{WB}. The parity of the measurements of the Shor state qubits is the outcome of the SM. The SM process thus costs four qubits per Shor state and an additional qubit for verification, for each of the six syndromes to be measured. This gives a total of 30 qubits versus the six needed for the single qubit SM. The complete Shor state SM is shown in Fig.~\ref{Shor}. Furthermore, to ensure there are no errors during the SM themselves the entire set of SM should be repeated until the same syndrome is read out twice. This raises the cost to a minimum of 60 qubits. 

Previous work has explore the frequency with which Shor stat SM should be applied during a sequence of logical gates. Reference \cite{How} explored analytically after how many logical gates Shor state SM should be implemented. The calculations suggested that with respect to accuracy there was in essence no difference between implementing SM after one, two, or three logical gatesthough clearly less resources are consumed if SM is implemented less often. This work was extended in \cite{How2} by numerically simulating the implementation of 50 logical gates on information encoded into a [[7,1,3]] QEC code, in different error environments. In the simulations Shor state SM were interspersed between the gates at various intervals to determine the optimum number of times SM should be implemented. The results demonstrated that, depending on the environment, it is not usually best to implement SM after every gate. In addition, the difference in fidelity between the optimum number of SM implementations and applying SM just once at the end of hte gate sequence was negligible. Hence, it may be more efficient to apply SM less often and save valuable resources.   

In this paper we extend the results of \cite{How2} by exploring two additional SM protocols. This will again be done by simulating 50 gates in different error environments with the alternate SM protocols interspersed between the gates at various intervals. We will then compare the performance of these schemes, as measured by accuracy and resource consumption, to that of the Shor state SM protocol. 

\begin{figure}
\begin{center}
\includegraphics[width=5cm]{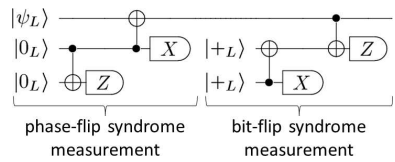}
\caption{Circuit for fault tolerant syndrome measurements for the [[7,1,3]] QEC code using Steane ancilla. Each line represents the seven physical qubits. The circuit shows that each ancilla is verified to check for errors that may have occurred during ancilla construction. }
\label{Steane}
\end{center}
\end{figure}

The first alternate SM protocol for the [[7,1,3]] QEC code that we simulate is via the use of Steane ancilla \cite{Steane} as shown in Fig.~\ref{Steane}. A Steane ancilla is a seven qubit system in the logical $|0\rangle$ or $|+\rangle$ state of the [[7,1,3]] QEC code. The states are utilized for the phase-flip and bit-flip SM, respectively. To identify a bit- (phase-) flip error a logical controlled-NOT (CNOT) gate is applied between the data qubits and ancilla qubits with the data qubits (ancilla qubits) as the control. The ancilla qubits are measured to determine the syndrome. To construct the ancilla in a fault tolerant fashion, two copies of the logical $|0\rangle$ or $|+\rangle$ state are constructed following the non-fault tolerant gate sequence of Ref.~\cite{Steane}. A logical CNOT gate is applied between the two copies and one is measured to check for possible construction errors in the other. In total we thus have a cost of 28 qubits to completely read out the syndromes. Unlike the Shor state method, Steane state SM is fault tolerant even if not repeated\cite{AGP}. Nevertheless, we will find that, at times, the output state fidelity will be higher if we do repeat the SM (at a total cost of 56 qubits). 

The second alternate SM protocol we explore is the single ancilla qubit protocol initially designed with the [[7,1,3]] QEC code. This protocol is not fault tolerant and thus presumably not scalable. Nevertheless, a non-fault tolerant protocol may still provide highly accurate results for problems of interest. In fact, recent work has begun to analyze different non-fault tolerant approaches to various quantum computing protocols with the goal of optimizing fidelity and resource consumption \cite{WB,YSW,BHW,YSWTgate,NFB,YT}.


\section{Simulation Model}

Our simulations allow us to determine the accuracy of implementing many logical gates with different SM protocols in a variety of error environments. The first step of our simulations is to perfectly encode a qubit of quantum information. We then start implementing logical gates in a noisy environment. Interspersed within the logical gates (at different intervals for different simulations) we peform SM (in the same noisy environment). 

Within the QFT framework the only proper gate operations are those which will keep quantum information protected at all times and are designed in such a way as to stem any possible spread of errors. For many QEC codes, including the [[7,1,3]] code, such gates include Clifford gates and the $T$-gate, a single-qubit $\pi/4$ phase rotation, which constitute a universal gate set. For Calderbank-Shor-Steane (CSS) codes, Clifford gates can be implemented bit-wise while the $T$-gate requires a specially prepared ancilla state and a series of controlled-NOT gates. 

Such a restrictive gate set means that many gates must be applied (and many resources consumed) to accomplish what may superficially appear to be a straightforward task. For example, much work has been done to determine optimal gate sequences for implementing an arbitrary single-qubit rotation (within prescribed accuracy $\epsilon$) using only the gates Clifford plus $T$ \cite{SK1,SK2,Svore1,KMM1,TMH,Svore2,KMM2,Selinger1,KMM3}. The goal of these works has been to find as short a gate sequence as possible with the fewest number of resource-consuming $T$-gates. Recent results allow, for example, a $\sigma_z$ rotation by 0.1 to be implemented within an accuracy of $10^{-5}$ by using 56 \cite{KMM3} $T$-gates, interspersed by one or two Hadamard, $H$, and Phase, $P$, gates. A fault tolerant implementation of this rotation would require more than 100 gates. Applying SM after each gate, as is assumed for fault tolerant quantum computation, thus costs thousands of qubits and hundreds of time steps just to implement a single rotation. 

In our simulations we perform 50 logical gates from the above set, 30 Clifford gates and 20 $T$-gates. These gates simulate fault tolerant performance of these gates as described in \cite{YSWTgate}. For a logical Clifford gate $C$ this means applying $C^{\dag}$ to each of the seven physical qubits. For the logical $T$-gate this requires constructing the ancilla state $|\Theta\rangle = \frac{1}{\sqrt{2}}(|0_L\rangle+e^{i\frac{\pi}{4}}|1_L\rangle)$, where $|0_L\rangle$ and $|1_L\rangle$ are the logical basis states of the [[7,1,3]] QEC code. A bit-wise CNOT gate is then applied between the state $|\Theta\rangle$ and the encoded state with the $|\Theta\rangle$ state qubits as control. The encoded state is measured, and if the outcome is zero the qubits of the $|\Theta\rangle$ state will be projected into the state $T$ acting on the encoded state. The construction, CNOT, and measurements are all peformed in the noisy environment.  

The error environment used in our simulations is the nonequiprobable Pauli operator error environment \cite{QCC} with non-correlated errors. As in \cite{AP}, this error model is a stochastic version of a biased noise model that can be formulated in terms of Hamiltonians coupling the system to an environment. In our simulations different error types arise with different probabilities. Individual qubits undergo $\sigma_x^j$ errors with probability $p_x$, $\sigma_y^j$ errors with probability $p_y$, and $\sigma_z^j$ errors with probability $p_z$, where $\sigma_i^j$, $i = x,y,z$ are the Pauli spin operators on qubit $j$. Only qubits taking part in a gate operation, initialization, or measurement will be subject to error while other qubits are perfectly stored. Qubits taking part in a two-qubit gate will undergo errors independently. The idealized assumption that idle qubits are not subject to error is partially justified in that it is expected such qubits will be less likely to undergo error (see for example \cite{Svore}). In our simulations we utilized four basic error environments: depolarization, $p = p_x = p_y = p_z$, and when one of the error probabilities, $p_i$ is dominant and the other two $p_j = p_k = 10^{-10}$. 
   
We assume a single qubit state $|\psi\rangle=\cos\alpha|0\rangle+e^{i\beta}\sin\alpha|1\rangle$, perfectly encoded into the [[7,1,3]] QEC code. The series of (necessarily noisy) logical gates, $U_{50}...U_2U_1$, is implemented in the nonequiprobable error environment with (noisy) SM interspersed amongst the gates leading to a final state, $\rho_f$, of the 7 qubits.  

We find it convenient to define composite gates $A = HPT$ and $B = HT$. These composite gates form the basic building blocks for gate sequences that implement arbitrary single qubit rotations from the gate set Clifford plus $T$. For our simulations we randomly chose a sequence of 20 composite gates (comprising 50 total gates):
\begin{equation}
U = ABBBAAAABBABABABBBAA,
\end{equation}
totalling 20 $T$ gates, 20 $H$ gates, and 10 $P$ gates. As menionted above, all physcial gates comprising the logical gates, measurements, and associated ancilla construction are simulated in the non-equiprobable error environment. 

To determine how often SM should be applied during a quantum computation we simulate the above 50 logical gates with SM performed at the following intervals: after every gate, ($q = 50$, where $q$ is the number of SM during the 50 logical gates), after every composite gate $A$ and $B$ ($q = 20$), after every two composite gates ($q = 10$), after every five composite gates ($q = 4$), after each half of the entire sequence ($q = 2$), and after the entire sequence ($q = 1$). Every physical gate of the SM implementation, including construction of the ancilla needed for the SM, is done in the nonequiprobable error environment. Each simulation is repeated for error environments of different values of $p_x$, $p_y$ and $p_z$. For all simulations, our initial state is the basis state $|0\rangle$. Other initial states and gate sequences were explored and give similar results.   

To determine the accuracy of the simulated implementations when compared with perfectly applied gates we utilize physical and logical state fidelities. The physical state fidelity is simply $F(\rho_i,\rho_f) = {\rm{Tr}}[\rho_i\rho_f]$, where $\rho_i$ is the ideal final state. To calculate the logical state fidelity we (perfectly) decode $\rho_f$ and trace out all qubits but the first. This leaves a one qubit state $\rho_f^{dec}$. The logical state fidelity is then $F(\rho_1,\rho_o^{dec}) = {\rm{Tr}}[\rho_1\rho_f^{dec}]$, where $\rho_1$ is the ideal single qubit final state. The former fidelity measure provides an accuracy measure for the entire evolutionary process of the physical qubits. It also informs us of the fragility of the physical state from which we can judge the harm a future error will do to the logical information. The latter fidelity measure is the accuracy of the encoded logical information. We will find it useful to utilize the physical and logical state infidelity $I = 1 - F$. 

Each operation applied in our simulations is done in a noisy error environment. Hence, the largest contribution to the non-unity of the final state fidelity will likely be dominated by the error of the final operation. While this is realistic, it hides information about the errors present in previous gates and how well they may have been corrected by SM and possible recovery operations. We thus apply perfect SM at the end of our entire sequence uncovering the affect of errors that are uncorrectable. We can then determine physical and logical state fidelities for the final state after perfect error correction. 

\begin{figure}
\begin{center}
\includegraphics[width=5.75cm]{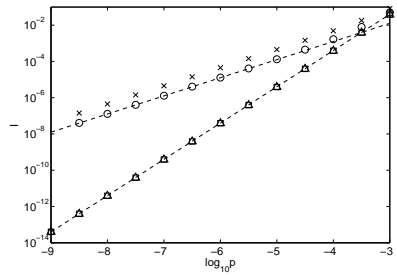}
\caption{Physical ($\times$) and logical ($\bigcirc$) infidelities for the output state after 50 logical gates with syndrome measurements applied after each gate, and physical ($\square$) and logical ($\triangle$) infidelities with perfect syndrome measurements appended as a function of depolarizing strength $p$. The dashed lines show infidelity proportional to $p$ and to $p^2$. }
\label{fids}
\end{center}
\end{figure}

Fig. \ref{fids} compares the four fidelity measures of the final state after 50 logical gates applied in a depolarizing environment with SM applied after each gate as a function of depolarizing strength, $p$. Not surprisingly, the fidelities without final perfect SM decrease proportionally to $p$, reflecting the error due to the final operation. The fidelities with final perfect syndromes decrease as $p^2$ as expected for a distance 3 QEC code. Note that the overall behavior for the physical and logical fidelities are very similar. This is because the errors are not in any way more or less concentrated on the code subspace than on the rest of the 7-qubit state. The similarity in behavior is true in all of our simulations. Thus, we will report only the logical fidelities.   


We present our results via a series of figures which allow us to determine the optimum syndrome measurement method as a function of both fidelity and resources consumed. We define the fractional change, $D$, of the infidelity when applying SM less often than after every gate as:
\begin{equation}
D(I_{50},I_{q}) = \frac{I_{50}-I_{q}}{I_{50}}
\end{equation}   
for $q = 20, 10, 4, 2, 1$. A positive fractional change means a higher fidelity when applying SM less often. Negative fractional change means the fidelity is higher when applying SM after every gate. However, it is important to note that applying SM after every gate is extremely resource intensive. Thus, even if it provides the highest fidelity, it may not be the optimal choice of SM application schemes.    

\section{Depolarization Environment}

We first analyze the infidelity of the output states after application of 50 gates and $q$ SM in a depolarizing environment. The five figures correspond to five different SM methods: the Shor state method which, as explained above, requires repetition of the SM to achieve fault tolerance, two variations of the Steane method, when SM are not repeated and when SM are repeated (though repetition of the SM are not necessary for fault tolerance), and two variations of the single qubit ancilla method, where SM are repeated and not repeated (neither of which are fault tolerant).

\begin{figure}
\begin{center}
\includegraphics[width=5.75cm]{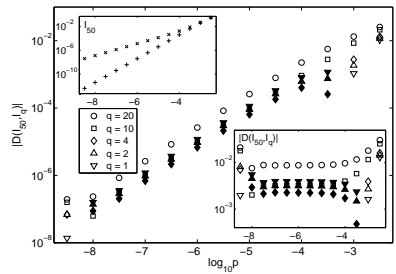}
\caption{Fractional change of logical infidelity for the final state after 50 logical gates as a function of depolarization strength for various syndrome measurement schemes $q$ with (lower right subplot) and without (main figure) perfect syndrome measurements applied at the end. Open shapes are for positive $D$ (applying syndromes after every gate gives worse fidelity), and filled shapes are for negative $D$ (applying syndromes after every gate gives better fidelity). The upper left subplot gives the logical infidelity as a function of depolarization strength with ($+$) and without ($\times$) a final perfect syndrome measurement. }
\label{ShorDepol}
\end{center}
\end{figure}

When implementing logical gates with Shor states in a depolarizing environment the best strategy for how often to apply SM can be determined from Figure \ref{ShorDepol}. For extremely low values of $p$ it is best to measure syndromes after every gate (not shown). This is hardly surprising and generally true since, at these low error probabilities, the cost of fidelity in applying SM is minimal. As $p$ increases the optimal choice of how often to apply SM requires the best balance between errors arising during gate implementation and errors arising from the SM themselves. Figure \ref{ShorDepol} shows that two schemes $q = 10, 20$ yield a higher fidelity than measuring syndromes after every gate with $q = 20$ giving the best fidelity. For $p \ge 10^{-3}$ measuring syndromes after every gate gives the lowest fidelity because the high fidelity cost of applying SM outweighs the gain in correcting gate implementation errors. 

An interesting feature of Figure \ref{ShorDepol} is the difference in behavior depending on whether or not final perfect syndrome measurements are applied. When they are not (main figure), the improvements in fidelity given by the best SM strategy increases linearly (on a logarithmic scale) with increasing depolarization strength. When they are (bottom-right inset) the improved fidelity is consistently between .01 and .001, except for large error strengths. This demonstrates that the perfect SM wipe out the affect of a final operation which reduces the fidelity by an amount proportional to $p$. Nonetheless, the choice of SM strategy giving the highest fidelity at any particular error strength remains the same whether or not perfect SM are appended.

\begin{figure}
\begin{center}
\includegraphics[width=5.75cm]{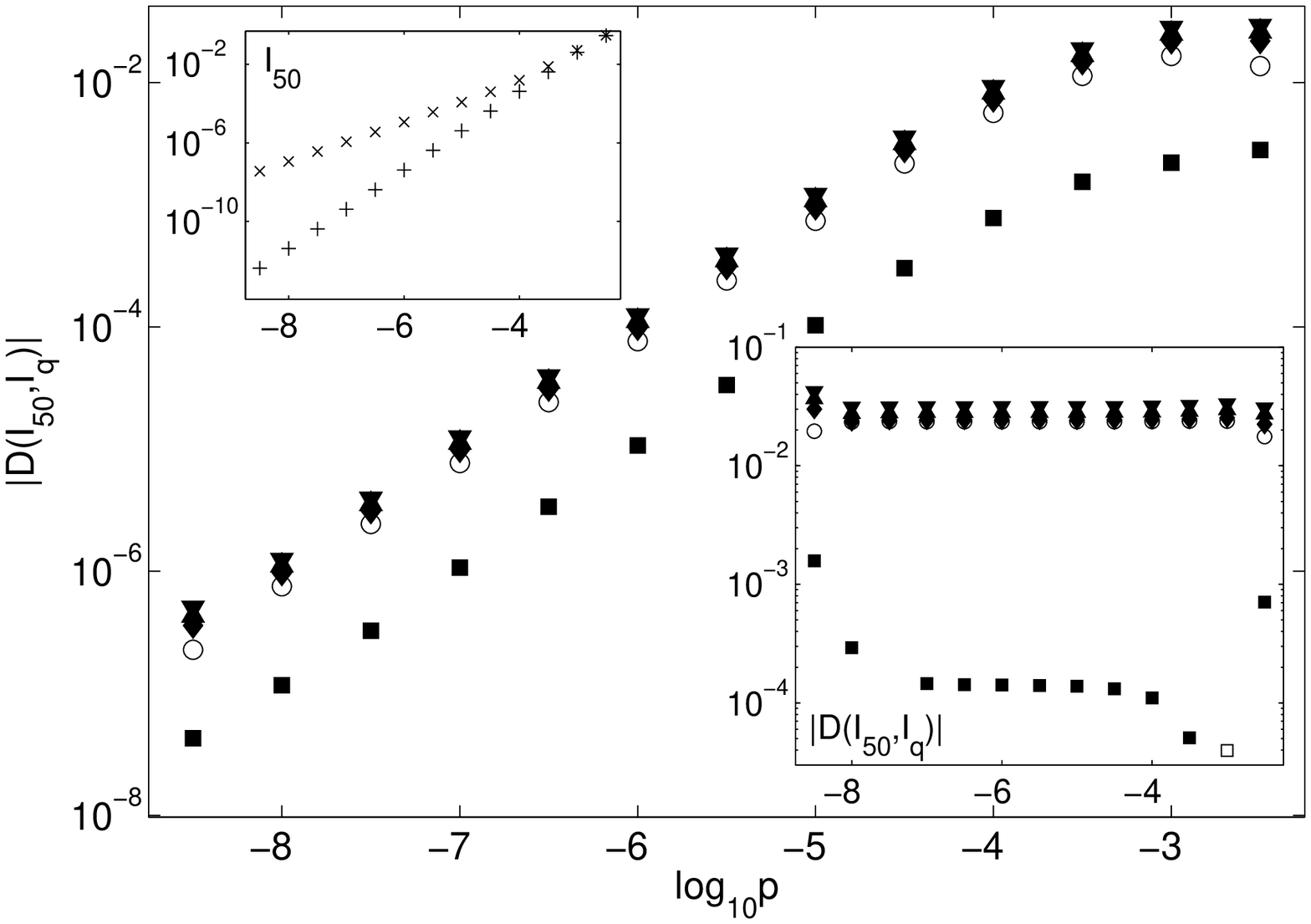}
\includegraphics[width=5.75cm]{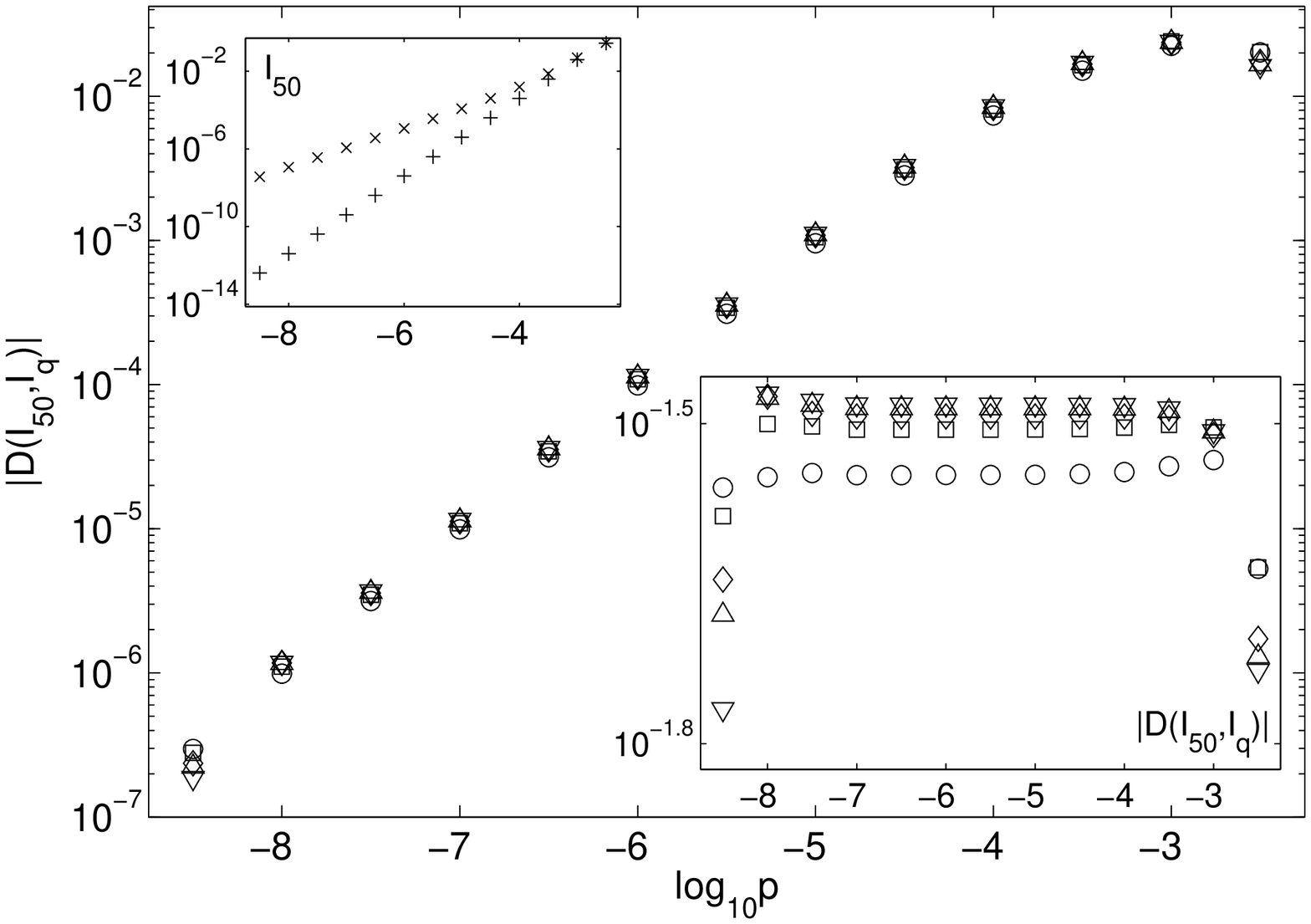}
\caption{Fractional change of logical infidelity for the final state after 50 logical gates as a function of depolarization strength, $p$, for various one-qubit ancilla SM schemes with (lower right subplot) and without (main figure) perfect syndrome measurements applied at the end. The SM are not repeated (left figure) or repeated (right figure). Open shapes are for positive $D$ (applying syndromes after every gate gives worse fidelity), and filled shapes are for negative $D$ (applying syndromes after every gate gives better fidelity). The upper left subplot gives the logical infidelity as a function of depolarization strength with ($+$) and without ($\times$) a final perfect syndrome measurement. }
\label{Depol1Q}
\end{center}
\end{figure}

\begin{figure}
\begin{center}
\includegraphics[width=5.75cm]{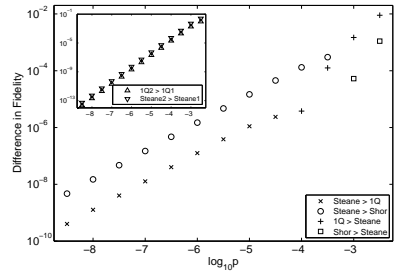}
\caption{Difference between the best fidelities (highest fidelity for all explored values of $q$ for each value of $p$) when using Steane anilla versus one-qubit ancilla and Shor state ancilla. For most of the range of error strengths we see that the Steane ancilla gives the best fidelity. However, at high error strength the single-qubit ancilla give the highest fidelity. The inset plots the difference between repeating and not repeating SM for the single qubit ancilla and Steane ancilla.}
\label{CompareDep}
\end{center}
\end{figure}

The one qubit syndrome ancilla is a non-fault tolerant SM method. Nevertheless, we see that the fidelity of the output state after 50 gates applying SM with the single qubit ancilla is actually slightly higher than the fidelity when SM are implemented with Shor states, as shown in Fig. \ref{CompareDep}. We look at two variations of this SM scheme: not repeating and repeating the SM. While the fidelity when applying SM after every gate is practically the same with both methods, we will see that different SM applications schemes lead to different results. 

When the SM are not repeated we find that the best SM scheme is $q = 20$ followed by applying SM after every gate as shown in Figure \ref{Depol1Q}. In contrast, when SM are repeated we find that in general the less times SM are applied the better, as shown in Figure \ref{Depol1Q}. For error probabilities $10^{-9} \le p \le 10^{-2}$ applying SM after every gate always gives the worst fidelity and applying SM once, $q = 1$ is generally the best. Thus, repeating SM leads to a factor of 10 savings in time and resources and, as shown in Figure \ref{CompareDep}, the fidelity of this scheme is slightly higher.  

\begin{figure}
\begin{center}
\includegraphics[width=5.75cm]{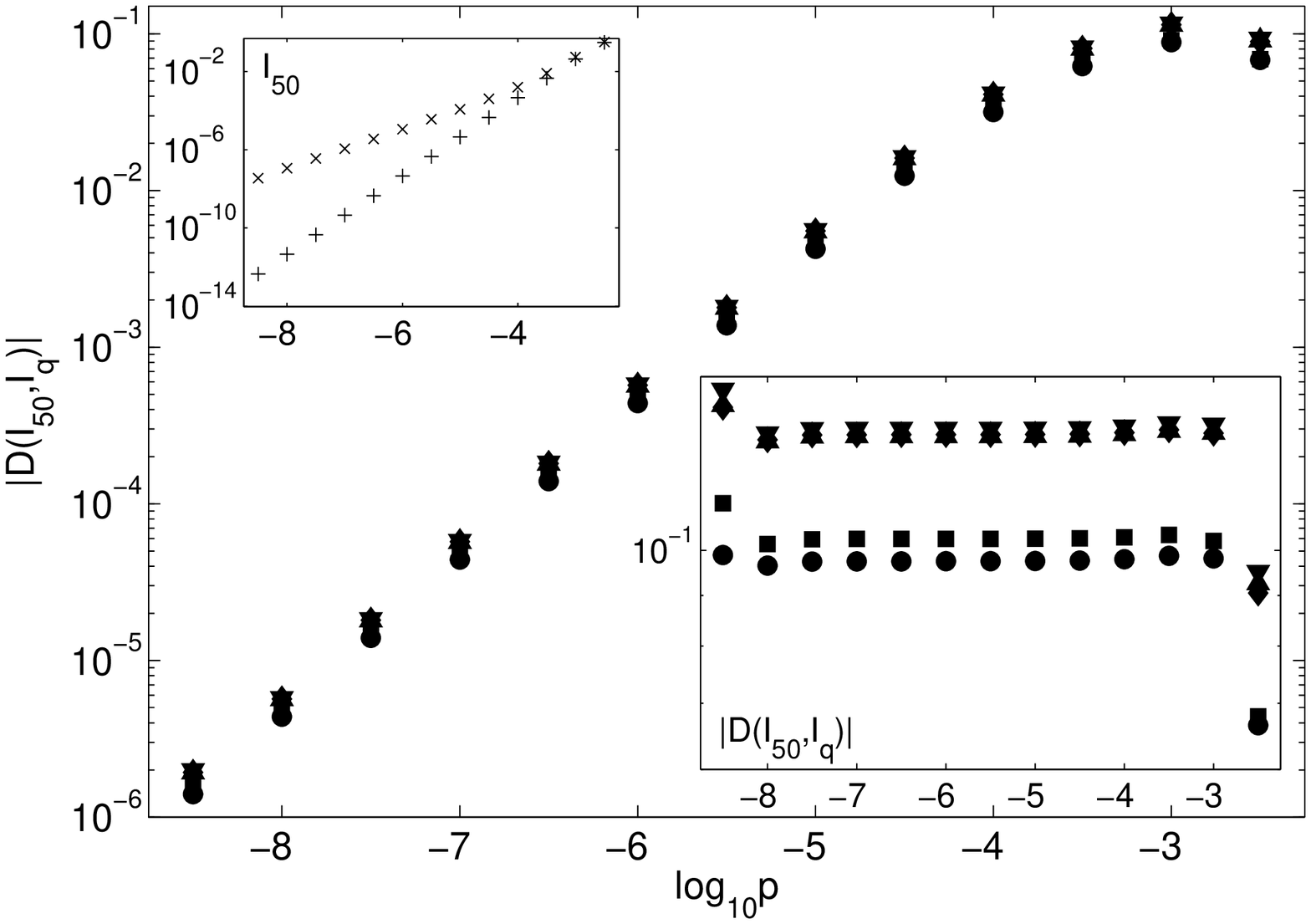}
\includegraphics[width=5.75cm]{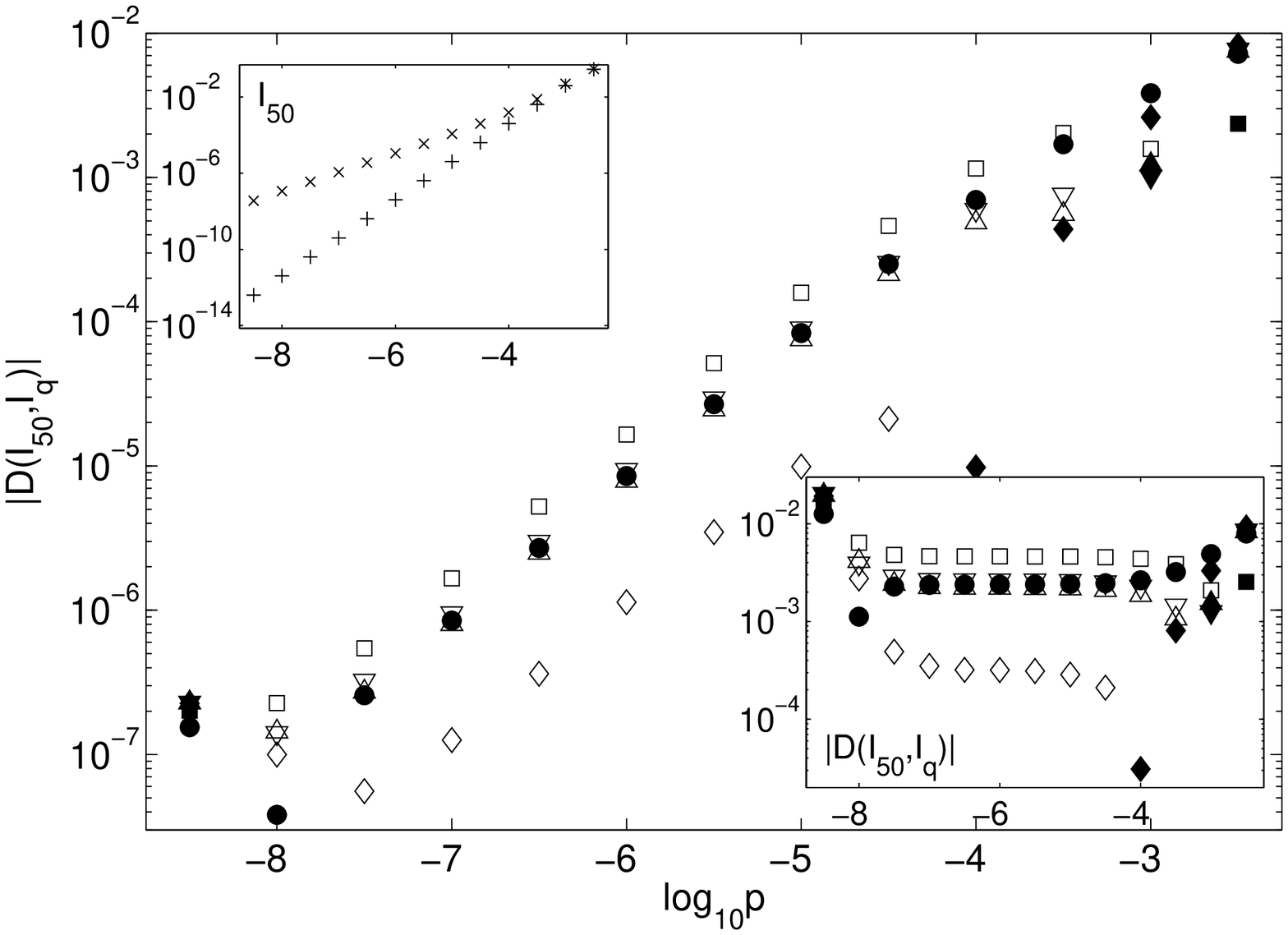}
\caption{Fractional change of logical infidelity for the final state after 50 logical gates as a function of depolarization strength for various Steane ancilla SM schemes with (lower right subplot) and without (main figure) perfect syndrome measurements applied at the end. For the SM each syndrome is measured once (left figure) or twice (right figure). Open shapes are for positive $D$ (applying syndromes after every gate gives worse fidelity), and filled shapes are for negative $D$ (applying syndromes after every gate gives better fidelity).The upper left subplot gives the logical infidelity as a function of depolarization strength with ($+$) and without ($\times$) a final perfect syndrome measurement. }
\label{DepolSteane}
\end{center}
\end{figure}

Using Steane states for SM gives the highest fidelity amongst the syndrome measurement methods explored here. When SM are not repeated (repetition is not required for fault tolerance) the best scheme is to apply SM after every gate as shown in Figure \ref{DepolSteane}. If SM are repeated the best scheme is $q = 10$ and the ordering of the other schemes depends on $p$ as shown in the Figure. Thus, repeating SM leads to a factor of 2.5 savings in time and resources and, in addition, the fidelity of this scheme is slightly better as shown in Figure \ref{CompareDep}.   

Figure \ref{CompareDep} compares the fidelity of the best of each of the three syndrome measurement strategies by plotting the difference between the Steane ancilla and the other two cases (fidelity of the output state using Steance ancilla minus fidelity of the output state using the other ancilla types). Clearly using Steane states for SM gives the highest fidelity for $p < 10^{-4}$ and the single qubit ancilla gives the hightest fidelity for higher error strengths. It should be noted, however, that the three fidelities are not that different and the single-qubit syndrome measurements gives a comparable fidelity despite the fact that it is not fault tolerant.  

\section{Asymmetric Error Environments}

The next set of results explore asymmetric error environments where one of the error probabilites $p_i$ is dominant and the other error probabilities $p_j, p_k$ are equal to $10^{-10}$. These environments demonstrate the effectiveness of tailoring SM to the particular evolution of the system as the approach to take depends strongly on which error is dominant and the dominant error strength. For all SM methods we find that the optimal $q$ depends strongly on the error environment.

When using Shor state for SM with dominant $\sigma_y$ errors the best fidelity is achieved when SM are applied after each gate when $p_y \le 10^{-5.5}$. For stronger (more asymmetric) errors  $q = 20$ gives higher fidelity. When phase-flip errors are dominant, applying SM after every gate gives the worst fidelity. The other schemes are about equal but the best fidelity is achieved for $q = 4$. When bit-flip errors are dominant we have already noted \cite{WB,How2} that not applying SM at all leads to the highest fidelity (not shown in figure). 

When not repeating one-qubit ancilla SM $q = 20$ gives the best fidelity when bit-flip errors are dominant, $q = 50$ gives the best fidelity when $\sigma_y$ errors are dominant, and $q = 4$ gives the best fidelity when phase-flip errors are dominant. However, when the one-qubit ancilla SM are repeated the situation changes dramatically. When bit-flip errors are dominant it is best to apply SM only after all 50 gates, $q = 1$. When $\sigma_y$ errors dominate $q = 50$ gives the best fidelity for error probabilities between $10^{-9}$ and $10^{-6}$. For higher error probabilities $q = 20$ gives the highest fidelity. When phase-flip errors are dominant, $q = 1$ again gives the highest fidelity.  

When using Steane states for SM we again see different and varying behavior. When SM are not repeated there is a sharp distinction between the environments when bit-flip errors dominate and when phase-flip errors dominate. When bit-flip errors dominate applying SM after every gate is optimal. When phase-flip errors dominate applying SM after all the gates is optimal. For $\sigma_y$ errors $q = 20$ gives the best fidelity. When SM are repeated it is best to apply SM only once during the 50 gates, $q = 1$ in the cases when $\sigma_y$ or phase-flips are the dominant error. If bit-flips are dominant the optimal amount of times to apply SM depends on the strength of $p_x$. If $p_x > 10^{-5.5}$ the best method is $q = 4$. For weaker $p_x$ the best method is $q = 50$. 

\begin{figure*}
\begin{center}
\includegraphics[width=5.75cm]{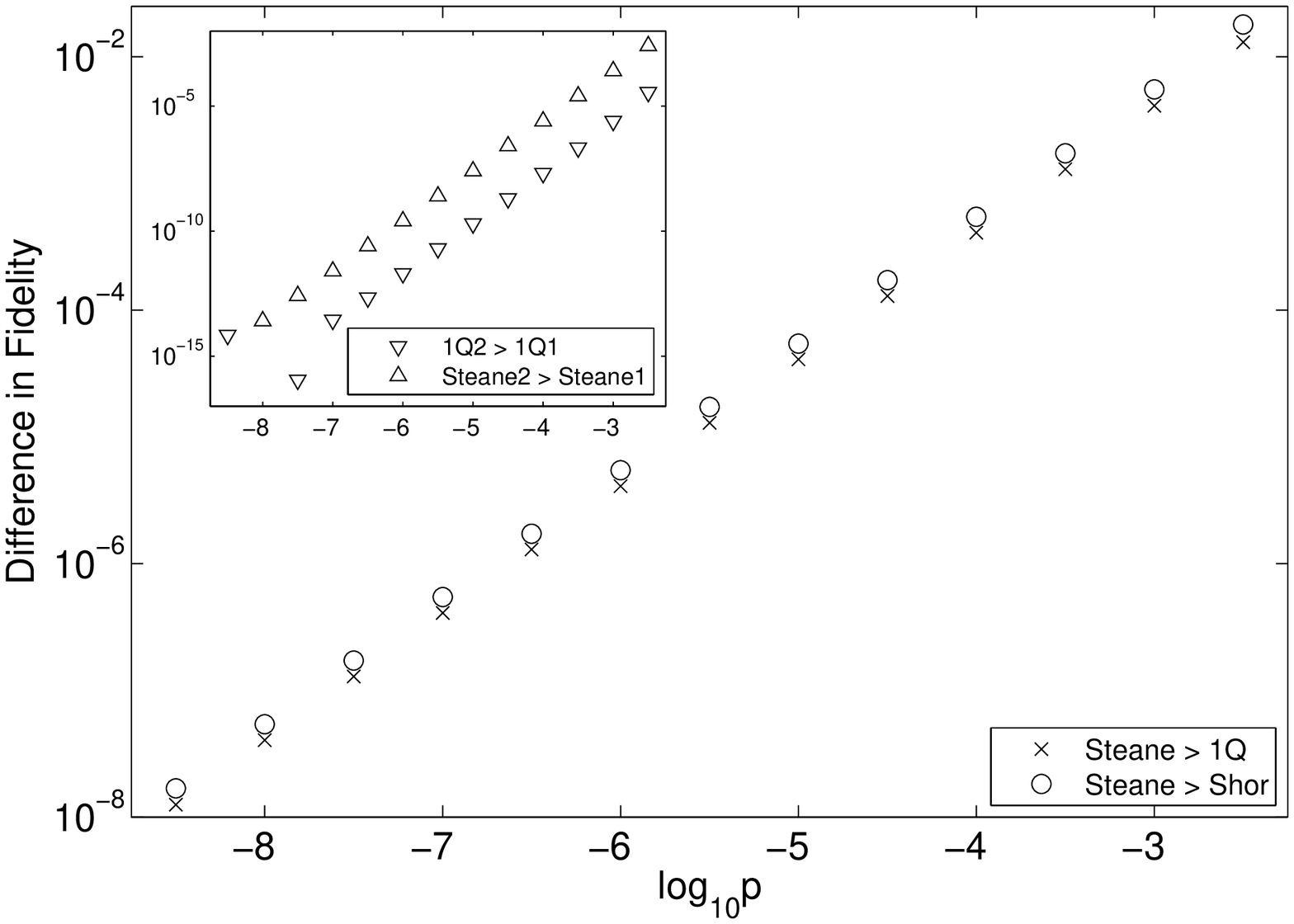}
\includegraphics[width=5.75cm]{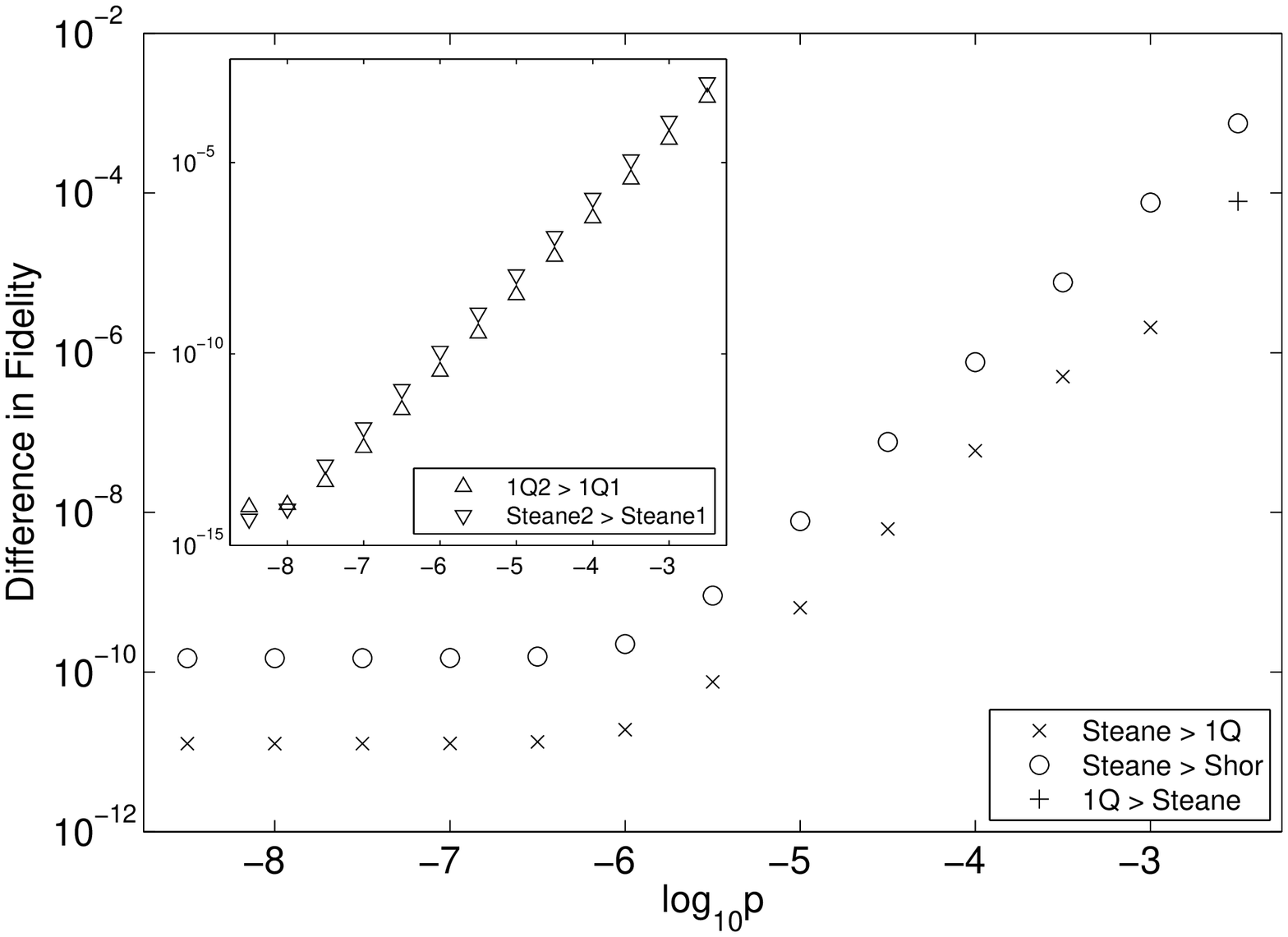}
\includegraphics[width=5.75cm]{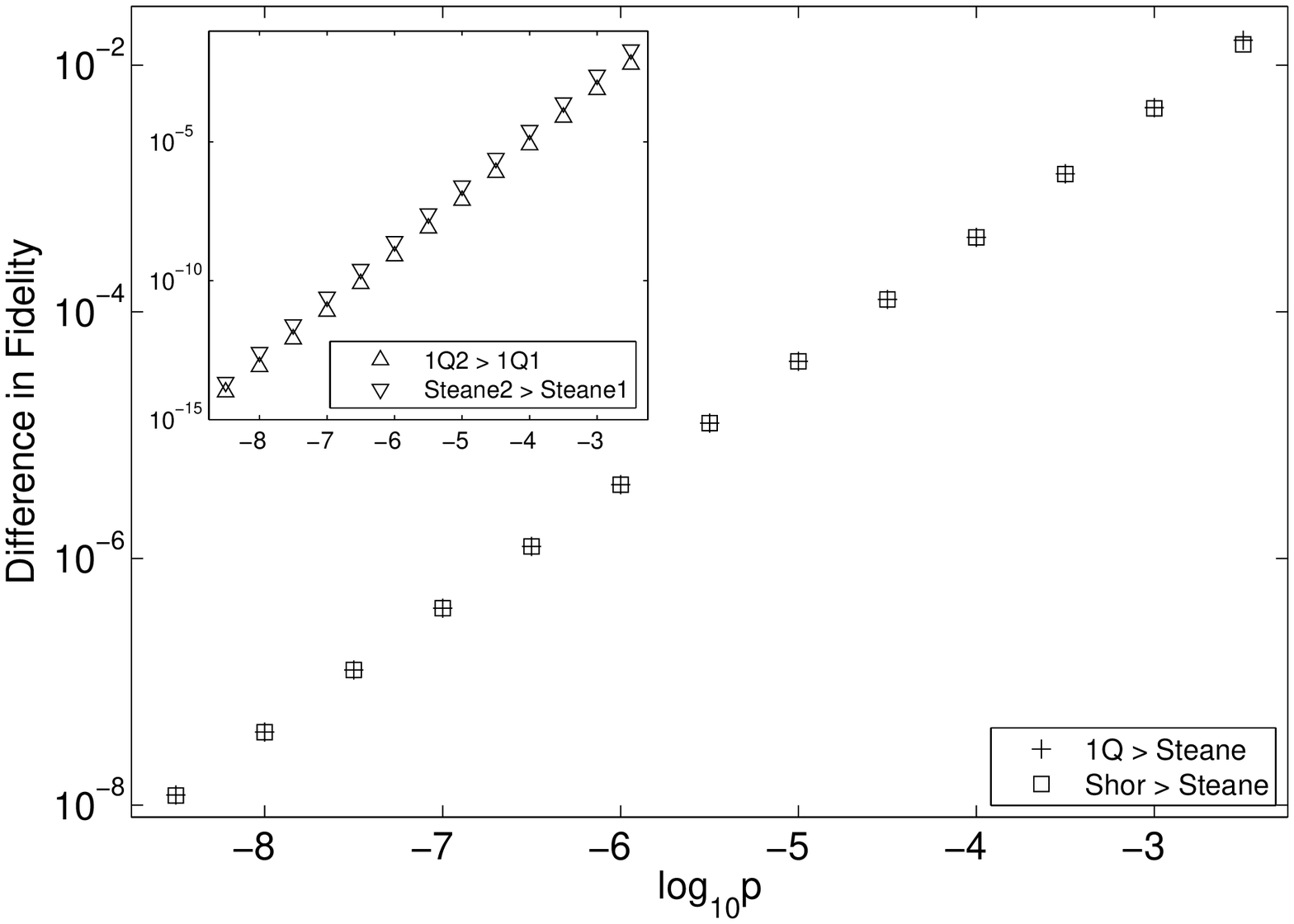}
\caption{Differences between the best fidelities (highest fidelity for all explored values of $q$ for each value of $p_j$) when using Steane anilla versus one-qubit ancilla and Shor state ancilla for different asymmetric error environments: dominant bit-flip errors, dominant $\sigma_y$ errors, and dominant phase-flip errors (bottom). In almost all cases the Steane ancilla gives the best fidelity. The insets plot the difference between repeating and not repeating SM for the single qubit ancilla and Steane ancilla.  
}
\label{CompareEnv}
\end{center}
\end{figure*}

Figure \ref{CompareEnv} compares the different SM methods for the asymmetric error models. The insets plot the difference in fidelity between repeating and not repeating SM for the single qubit and Steane ancillas. In all cases repeating SM leads to higher fidelity. In the case of the single qubit ancillas and bit-flip dominated environments the higher fidelity also comes with a factor of 10 savings in resources since $q = 1$ gives the best fidelity as opposed to $q = 20$ when the SM are not repeated. When $\sigma_y$ errors are dominant attaining the highest fidelity requires the use of twice as many resources for low $p$ but slightly less resources (40 SM as opposed  to 50) for higher $p$. When phase-flip errors dominate we use only half the resources when repeating SM.

Comparing the two Steane state SM variations, for a bit-flip dominated environment we see that to achieve the highest fidelity requires repeating SM at a cost of twice the resources for low error strengths but at a savings of more than a factor of 6 (8 SM versus 50SM) for higher error strengths.  When $\sigma_y$ errors dominate applying SM twice gives a slightly higher fidelity and a factor of 10 in resource savings. Finally, when phase-flip errors dominate applying SM twice per QEC gives a slightly higher fidelity but at a cost of two times the resources.  

Figure \ref{CompareEnv} compares the fidelity of the best of each of the three syndrome measurement methods in asymmetric error environments. For the bit-flip and $\sigma_y$ error dominated environments the Steane state ancilla give the best fidelity, by a wide margin for bit-flips but by a much smaller margin for $\sigma_y$ errors. In a phase-flip dominated environment the Steane syndrome measurements give the lowest fidelity. In all three error environments the fidelity of the Shor state syndromes and that of the single-qubit syndromes are similar, though the single qubit syndromes give a slightly better fidelity.    

\section{Conclusion}

In conclusion we have explored different syndrome measurement techniques for the [[7,1,3]] QEC code in different error environments. The three SM types explored are those using Shor states and Steane states, both of which are fault tolerant, and using a single qubit ancilla, which is not fault tolerant. We explore two variations of both the Steane state and single-qubit SM methods: the SM are repeated or not repeated. The simulations implement 50 logical gates within the [[7,1,3]] QEC encoding with SM applied at various intervals using the different types of SM. The results demonstrate that the choice of SM leading to the highest fidelity, and the number of times SM should be applied during the 50 gates that will optimize fidelity, will strongly depend on the error environment. In a depolarizing environment and when bit-flip and $\sigma_y$ errors dominate, the SM type leading to the higest fidelity is the Steane states. Even so, the number of times the SM should be applied to give this highest fidelity is different for the three environments. When phase-flip errors are dominant the Steane state SM give the lowest fidelity. The Shor state and single-qubit SM always give similar fidelities with the single-qubit SM fidelity being slightly higher. 

The different SM methods utilize different amount of resources and our results allow us to properly weigh resource consumption, time and the number of qubits, versus accuracy as measured via fidelity. While a given SM may provide the highest overall fidelity the cost in resources may prove overburdensome. For example, when performing the 50 gates in a depolarizing environment the highest fidelity is achieved using Steane ancilla applied every five logical gates ($q = 10$) and repeating the SM at a total cost of 560 qubits. Applying repeating one-qubit SM just once after all 50 gates costs only 2 qubits with a sacrifice in fidelity of about $p/10$. 

Furthermore, we note that despite the fact that the single-qubit SM are not fault tolerant, they generally achieve a fidelity that is relatively close to the other SM, at least over 50 gates. Whether this condition will hold over larger numbers of gates will be explored elsewhere but it will likely depend strongly on the details of the error environment and how often SM are applied. Nevertheless, these simulations suggest that at least for small computations high accuracy may be achievable without strictly following all the tenets of QFT.    

I would like to thank G. Gilbert for insightful comments. This research is supported under MITRE Innovation Program Grant 51MSR662. 

\end{document}